   \documentclass[preprint,showpacs,preprintnumbers,amsmath,amssymb,superscriptaddress]{revtex4}
\usepackage{graphicx} 
\usepackage{dcolumn}  
\usepackage{bm}       

\begin{document}

\preprint{To be submitted to PRC}

\title{Pair breaking and Coulomb effects in cold fission from reactions $^{233}$U(n$_{th}$,f), $^{235}$U(n$_{th}$,f) and $^{239}$Pu(n$_{th}$,f)}

\author{Modesto Montoya}
\email{mmontoya@ipen.gob.pe}

 \affiliation{Instituto Peruano de Energ\'ia Nuclear, Canad\'a 1470, San Borja, Lima, Per\'u.}
\affiliation{Facultad de Ciencias, Universidad Nacional de Ingenier\'ia, Av. T\'upac Amaru 210, R\'imac, Lima, Per\'u.}

\date{\today} 

\begin{abstract}
 In this paper, the distribution of mass and kinetic energy in the cold region of the thermal neutron induced fission of $^{233}$U, $^{235}$U and $^{239}$Pu, respectively, is interpreted in terms of nucleon pair-breaking and the Coulomb interaction energy between complementary fragments (“Coulomb effect”). The fission process ends in the scission configuration, in which complementary fragments 1 and 2, having proton numbers $Z_{1}$ and $Z_{2}$, and mass number $A_{1}$ and $A_{2}$, only interact each other with electrostatic repulsion,  by what each one acquires kinetic energy $E_{1}$ and $E_{2}$, respectively. However, before reach detectors, they emit $n_{1}$ and $n_{2}$ neutrons. Thus, the final detected fragments have mass numbers $m_{1}$ ($=A_{1}-n_{1}$) and $m_{2}$ ($=A_{2}-n_{2}$), and kinetic energies $e_{1}$ ($=E_{1}[1-n_{1}/A_{1}]$) and $e_{2}$ ($=E_{2}[1-n_{2}/A_{2}]$), respectively.  In order to avoid the erosive consequences of neutron emission, one studies the cold fission regions, corresponding to total kinetic energy ($TE$) close to the maximum available energy of the reaction ($Q$). Contrary to expected, in cold fission is not observed high odd-even effect in mass number distribution. Nevertheless, the measured values are compatible with higher odd-even effects on proton or neutron number distribution, respectively. In addition, in cold fission, the minimal total excitation energy ($X$) is correlated with the "Coulomb energy excess" ($\delta C$), which is defined as the difference between $C$ (the electrostatic interaction energy between complementary fragments taken as spherical in scission configuration) and $Q$.  These “Coulomb effects” increase with the asymmetry of the charge fragmentations. In sum, the experimental data on cold fission suggest that scission configurations explore all the possibilities permitted by the available energy for fission.
\end{abstract}


 \maketitle

\section{Introduction}
\begin{figure}
\centering
\includegraphics[width=8cm]{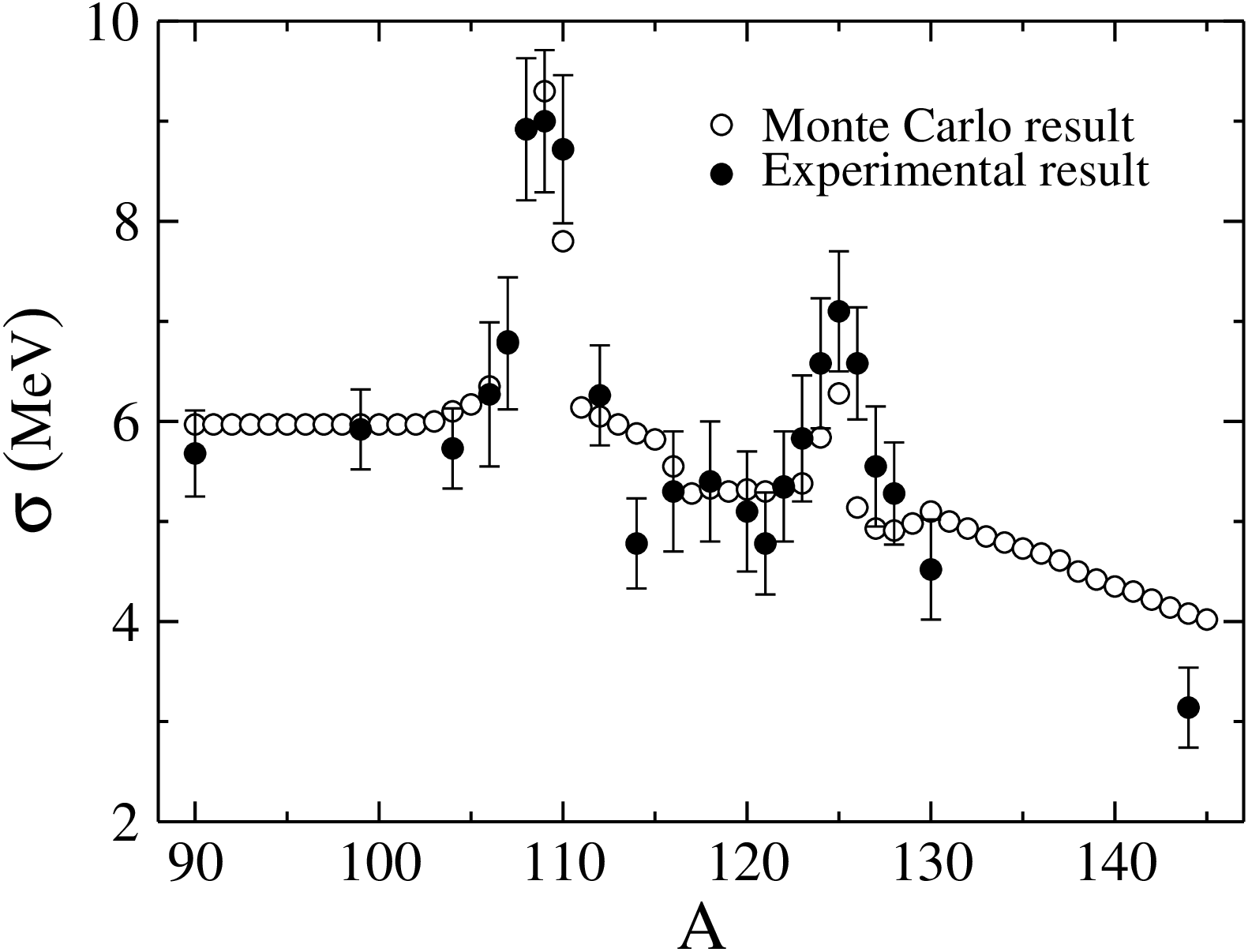}
\caption{Belhafaf \textit{et. al.} experimental \cite{Belhafaf83} and Monte Carlo simulation \cite{Montoya07,Montoya08} results on standard deviation of the kinetic energy distribution as a function of fragment mass number in the reaction $^{235}$U(n$_{th}$,f). See text.}
\label{fig1}
\end{figure}

Among the most studied properties of the nuclear fission of actinides are the distributions of mass, charge and kinetic energy associated to complementary fragments. However, R. Brissot et al. \cite{Brissot79} observed that the neutron emission from the fragments, occurred before they reach the detector, generates a structure on the final kinetic energy distribution as a function of mass, non-existent on primary fragment distribution. Naturally, neutron emission also erodes odd-even effects on neutron number distribution and consequently on mass number distribution associated to fragments. For these reasons, cold fission regions, corresponding to high kinetic energy associated to fragments that do not emit neutrons, are studied. In that sense, in this paper, mass, charge and kinetic energy distribution in the cold regions from thermal neutron induced fission of $^{233}$U, $^{235}$U and $^{239}$Pu, namely $^{233}$U(n$_{th}$,f), $^{235}$U(n$_{th}$,f) and $^{239}$Pu(n$_{th}$,f), respectively, is interpreted in terms of nucleon pair breaking and Coulomb interaction energy between fragments in scission configuration.\\\\
Let us suppose that the complementary fragments have kinetic energies $E_1$ and $E_2$, referred to primary mass numbers $A_1$ and $A_2$, and proton numbers $Z_1$ and $Z_2$, respectively. After the neutron emission, complementary fragments 1 and 2 end with mass numbers
\begin{equation*}
 m_{1}=A_{1}-n_{1}
\end{equation*}
and
\begin{equation*}
 m_{2}=A_{2}-n_{2}
\end{equation*}
where $n_1$ and $n_2$ are the numbers of neutrons emitted by the fragments 1 and 2, respectively. The corresponding values of the final kinetic energy associated to those fragments are
\begin{equation*}
e_1 \cong E_1 \left(1- \dfrac{n_1}{A_1} \right),
\end{equation*}
and
\begin{equation*}
e_2 \cong E_2 \left(1- \dfrac{n_2}{A_2} \right).
\end{equation*}
In the case that only one of the complementary fragments is detected, which is what happens in the Lohengrin spectrometer at the High Flux Reactor (HFR) of the Laue-Langevin Institute (ILL) in Grenoble \cite{Brissot79}, subscripts are omitted. For a given primary mass number ($A$), the kinetic energy ($E$) distribution is characterized by the corresponding average, $\overline{E}(A)$, and standard deviation, $\sigma_{E}(A)$.\\\\
Using the Lohengrin spectrometer in the HFR of ILL in Grenoble, R. Brissot \textit{et. al.} measured $\overline{e}(m)$ y $\sigma_{e}(m)$ values in $^{235}$U($n_{th}$,f). Their experimental curve $\sigma_{e}(m)$ showed a peak around $m=110$, as it had been predicted by a Monte Carlo simulation performed before that the experiment was carried out. In this simulation it was demonstrated that that broadening $e(m)$ distribution is due to the neutron emission and the sharp fall on $\overline{E}(A)$ in the mass region around $m=110$. For each mass number $m$,the kinetic energy distribution is the result of superposition of distributions associated to fragments having $A \geq m$ \cite{Brissot79}.\\\\
In the final mass region $106\leq m \leq 112$, $\overline{e}$ falls sharply, about 4 MeV per mass number. Let us suppose that the $\overline{E}$ behaviour is similar to that of $\overline{e}$, and $\sigma_{E}(A)$ is flat (equal to 5 MeV, for example). Therefore, the average kinetic energies associated to fragments with $A=112$ is 8 MeV lower than that of fragments associated to $A=110$. Fragments with $A=110$, that do not emit neutrons, have the high values of their distribution of primary energies $(E)$; and fragments associated to $A= 112$, that do emit 2 neutrons, have the low $E$ values of their corresponding primary distribution, whose average is already 8 MeV lower than that associated to $A=110$. Both fragments, one with primary mass numbers 110 y and the other with 112, respectively, having $E$ values with average differences of around 8 MeV, end with $m=110$. The superposition of these dispersed values contributes with the widening of $\sigma_{e}(m)=110$.\\\\
D. Belhafaf \textit{et. al.}, repeating the experiment on $^{235}$U(n$_{th}$,f) made by R. Brissot \textit{et. al.} \cite{Brissot79}, observed a broadening of the distribution of $e$ around $m=126$ \cite{Belhafaf83}. Moreover, for $^{233}$U(n$_{th}$,f), those authors observe a broadening of the distribution of $e$ around $m=124$. Their simulation did not reproduce these experimental results \cite{Belhafaf83}.\\\\
For $^{233}$U(n$_{th}$,f) H. Faust \textit{et. al.} \cite{Faust04}, using theoretical models on fission dynamics, calculated the distribution of $E$ as a function of $A$, but not around $A=124$. However the D. Belhafaf \textit{et. al.} experimental results for both, $^{233}$U(n$_{th}$,f) and $^{235}$U(n$_{th}$,f), were reproduced by a Monte Carlo simulation performed by de M. Montoya \textit{et. al.} \cite{Montoya07,Montoya08}. These authors shows that the broadening on the distribution of $e$ is generated by the superposition of the distributions of $E$ corresponding to $A\geq m$. For the case of $^{235}$U(n$_{th}$,f), $\overline{e}$ falls from 88 MeV referred to $m=125$, to 84 MeV referred to $m= 126$, while in this region the tendency of $\overline{e}$ is to increase with $m$ \cite{Belhafaf83}. See Fig. \ref{fig1}. It is reasonable to assume that for $A$ near but higher than 126, $\overline{E}$ has a similar behavior. Thus, the superposition $E$ values, corresponding to primary mass numbers $A \geq 126$, having dispersive behavior, contribute to the broadening of $e(m)$ distribution around $m=126$.

\section{Neutron emission and odd-even effects in cold fission}
\begin{figure}
\centering
\includegraphics[width=8cm]{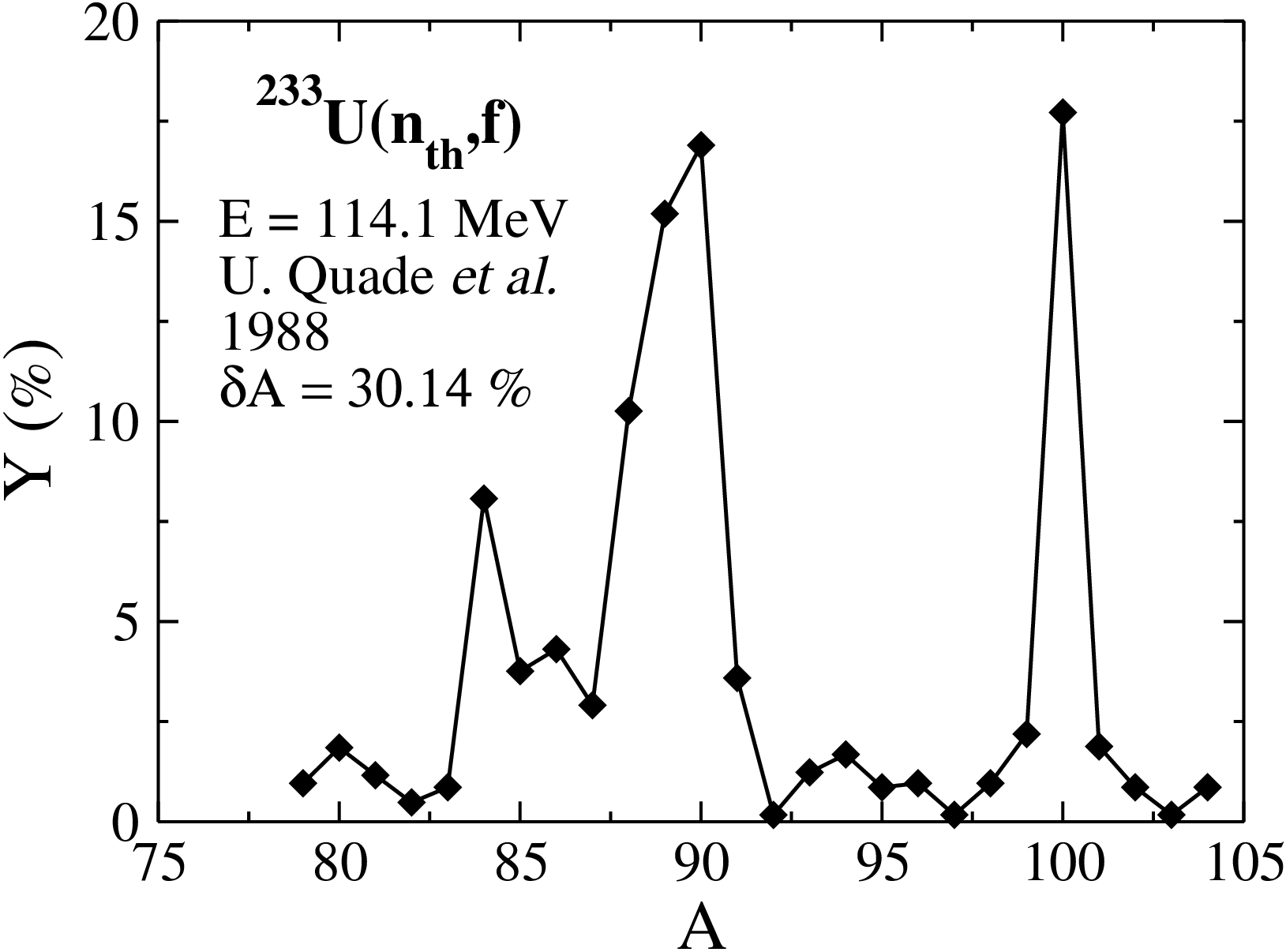}
\caption{Mass number distribution in the energy window $E=$ 114.1 MeV for cold fission from $^{233}$U(n$_{th}$,f). Taken from Ref. [\onlinecite{Clerc86}]. From these data one calculates $\delta A$= 30.14\%. See text.}
\label{fig2}
\end{figure}

\begin{figure}
\centering
\includegraphics[width=8cm]{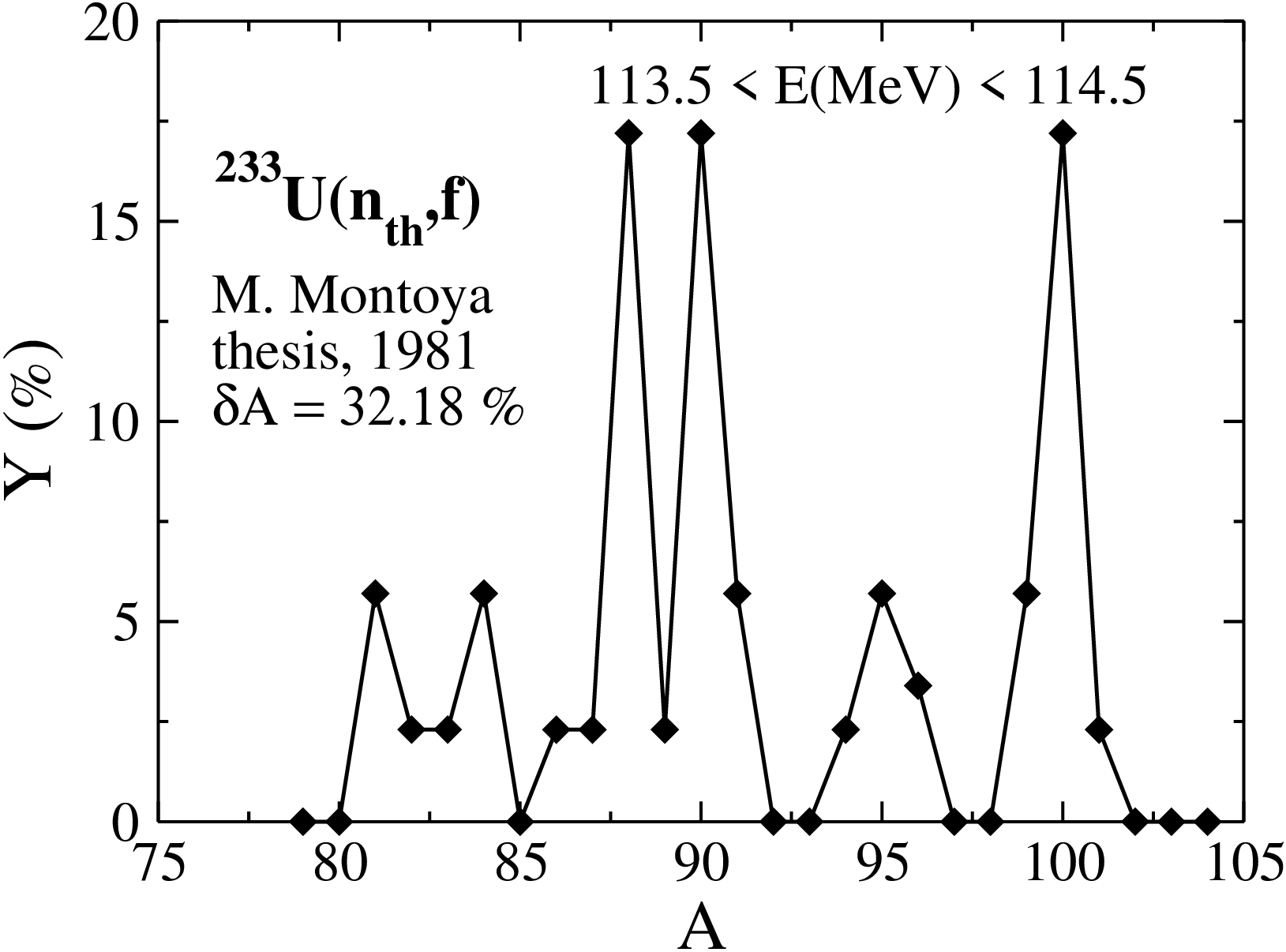}
\caption{Mass number distribution in the energy window 113.5 $<E<$ 114.5 MeV for cold fission from  $^{233}$U(n$_{th}$,f). Taken from Ref. [\onlinecite{Montoya81}]. From these data one calculates $\delta A=$ 32.18\%. See text.}
\label{fig3}
\end{figure}

\begin{figure}
 \centering
 \includegraphics[width=8cm]{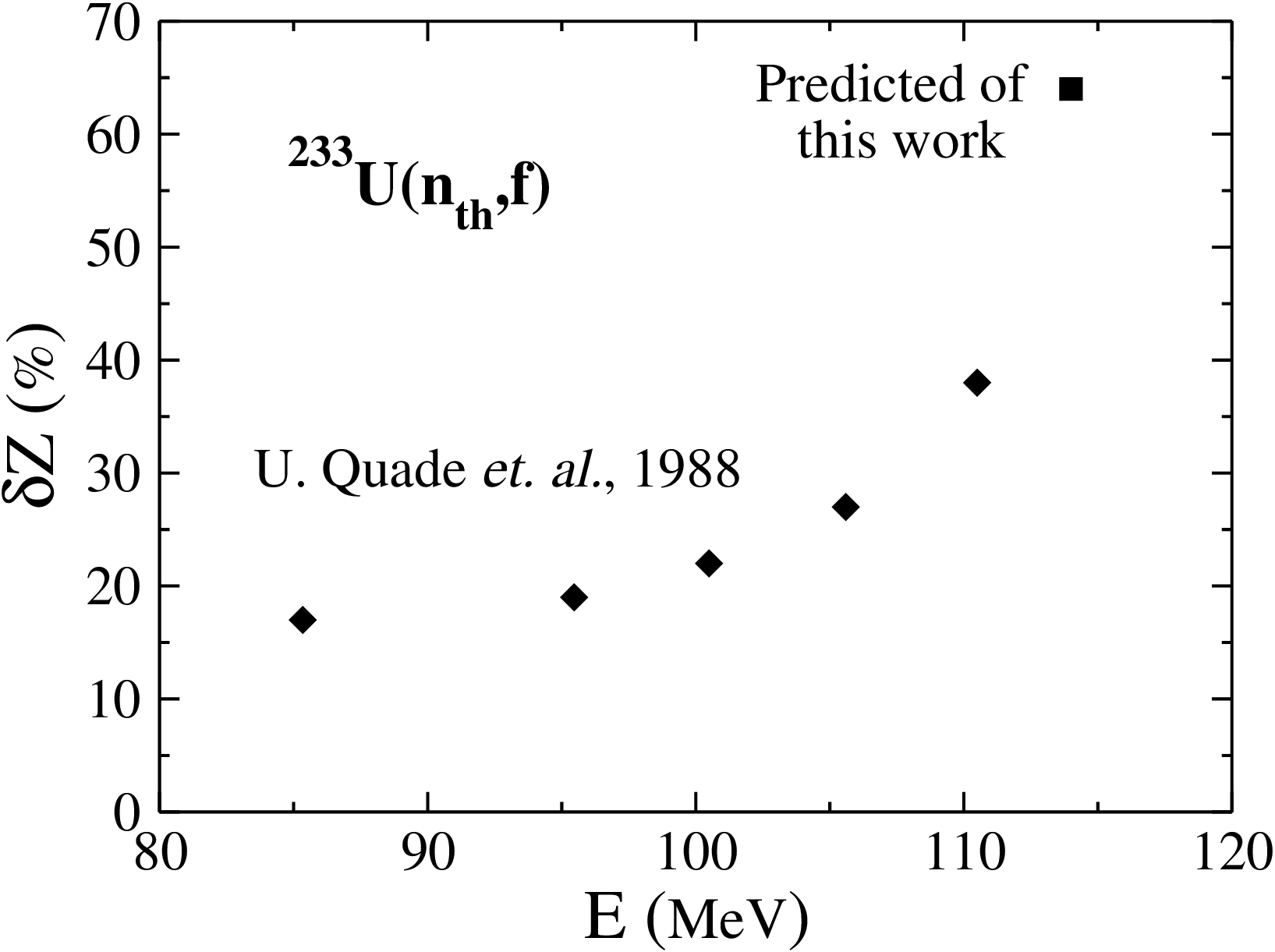}
 \caption{Odd-even effects on the proton distribution as a function of $E$ in $^{233}$U(n$_{th}$,f). Values are taken from Ref. [\onlinecite{Quade88}], except the higher one, which corresponds to prediction of this work. See text.}
 \label{fig4}
 \end{figure}

\begin{figure}
 \centering
 \includegraphics[width=8cm]{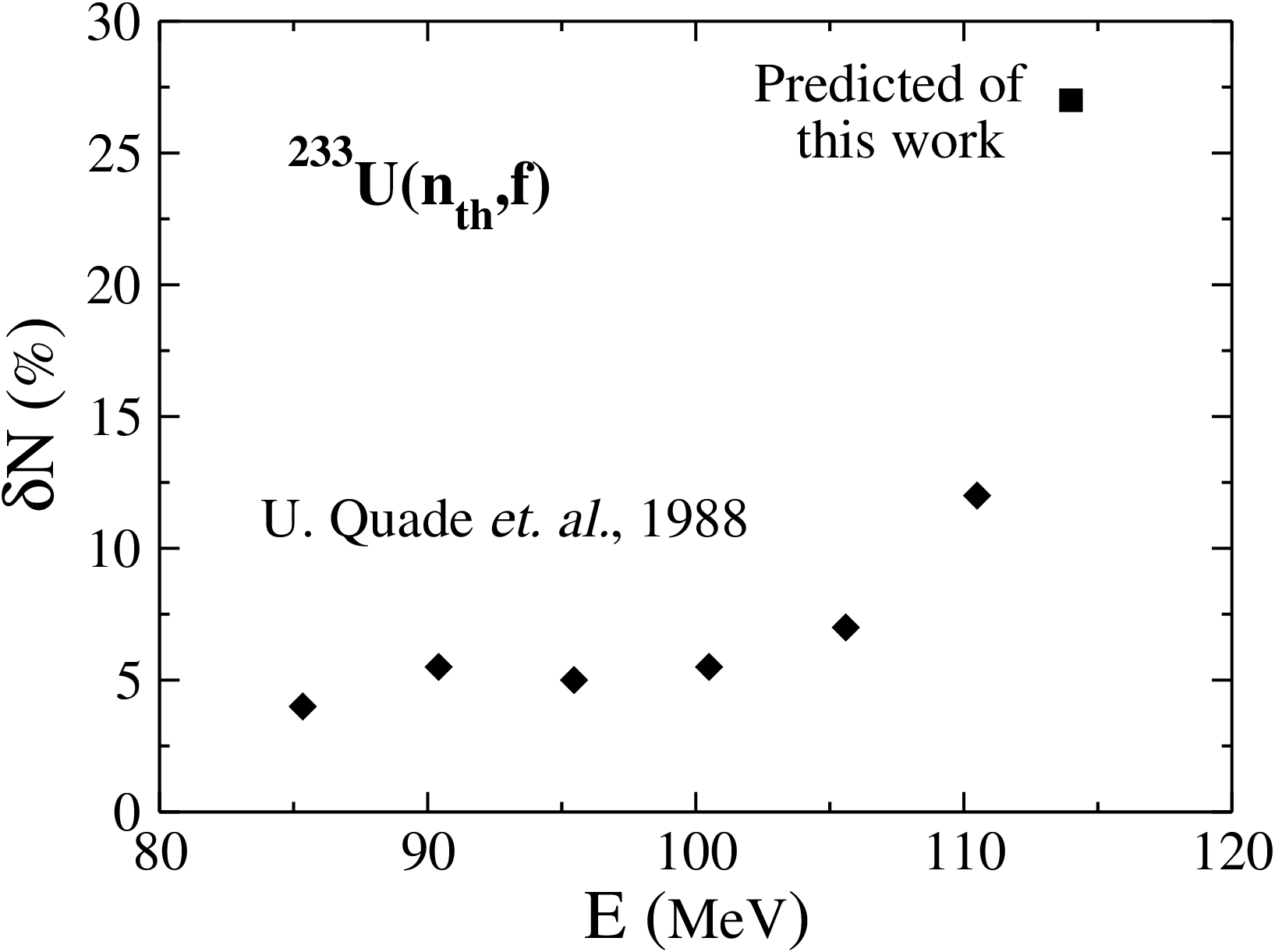}
 \caption{Odd-even effects on the neutron distribution as a function of $E$ in $^{233}$U(n$_{th}$,f). Values are taken from Ref. [\onlinecite{Quade88}], except the higher one, which corresponds to prediction of this work. See text.}
 \label{fig5}
 \end{figure}

\begin{figure}
 \centering
 \includegraphics[width=8cm]{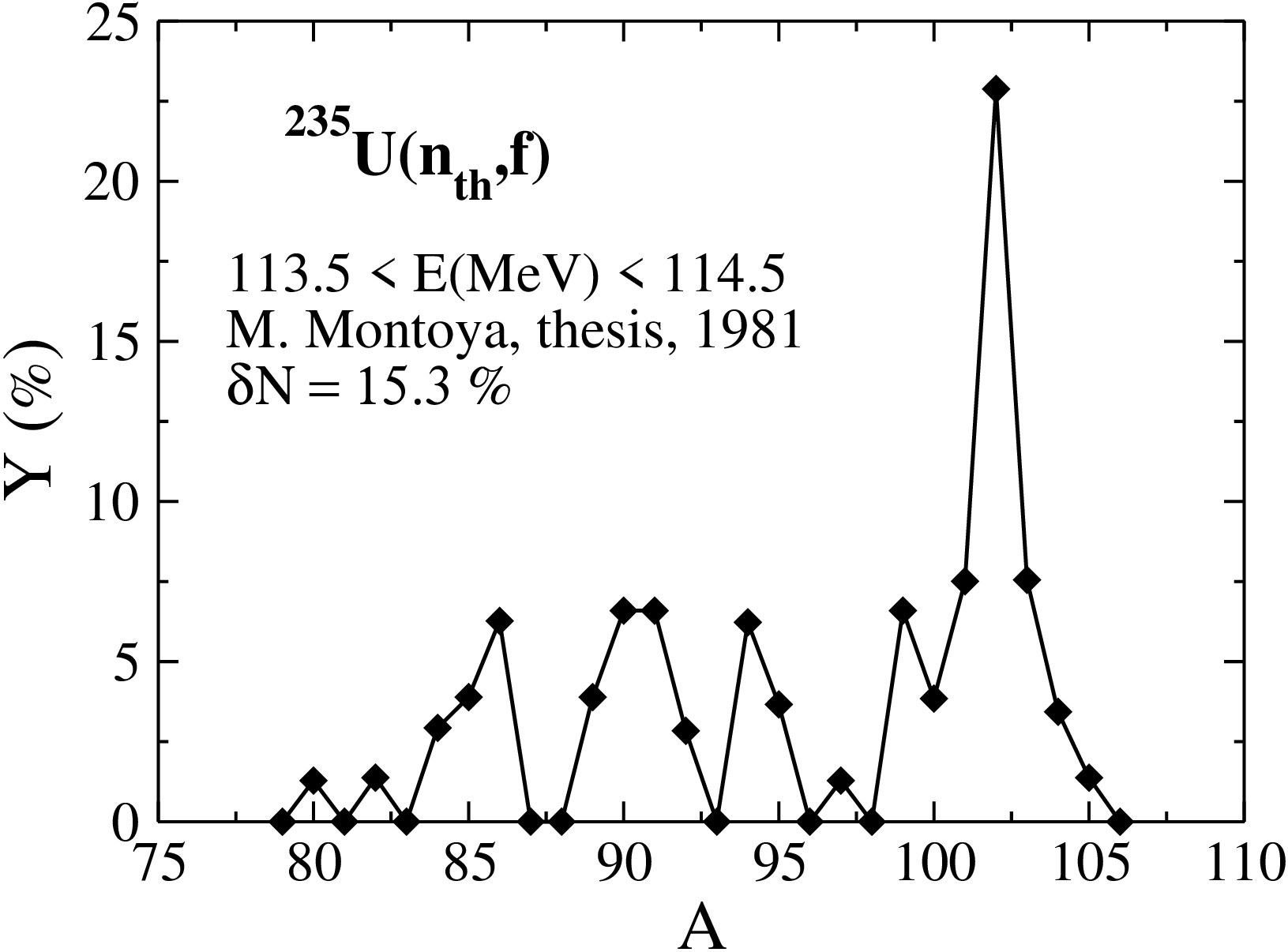}
 \caption{Mass number distribution in the energy window 113.5 $<E<$ 114.5 MeV for cold fission from $^{235}$U(n$_{th}$,f). Taken from Ref. [\onlinecite{Montoya81}]. From these data one calculates $\delta A$= 15.3\%. See text.}
 \label{fig6}
 \end{figure}

\begin{figure}
 \centering
 \includegraphics[width=7cm]{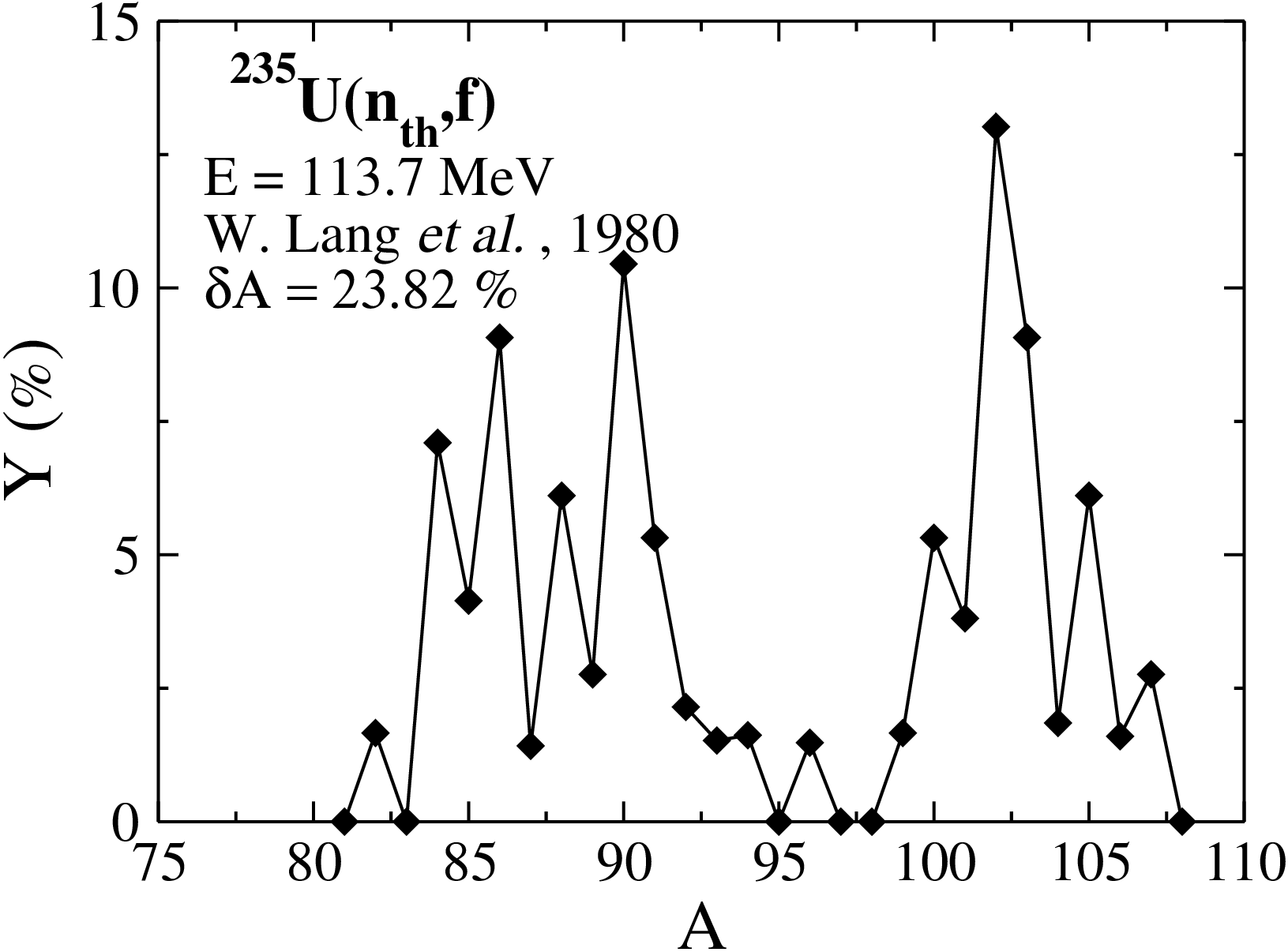}
 \caption{Mass number distribution in the energy window $E=$ 114.5 MeV for cold fission from the reaction  $^{235}$U(n$_{th}$,f). Taken from Ref. [\onlinecite{Lang80}]. From these data one calculates $\delta A$= 23.82\%. See text.}
 \label{fig7}
 \end{figure}

\begin{figure}
\centering
 \includegraphics[width=8cm]{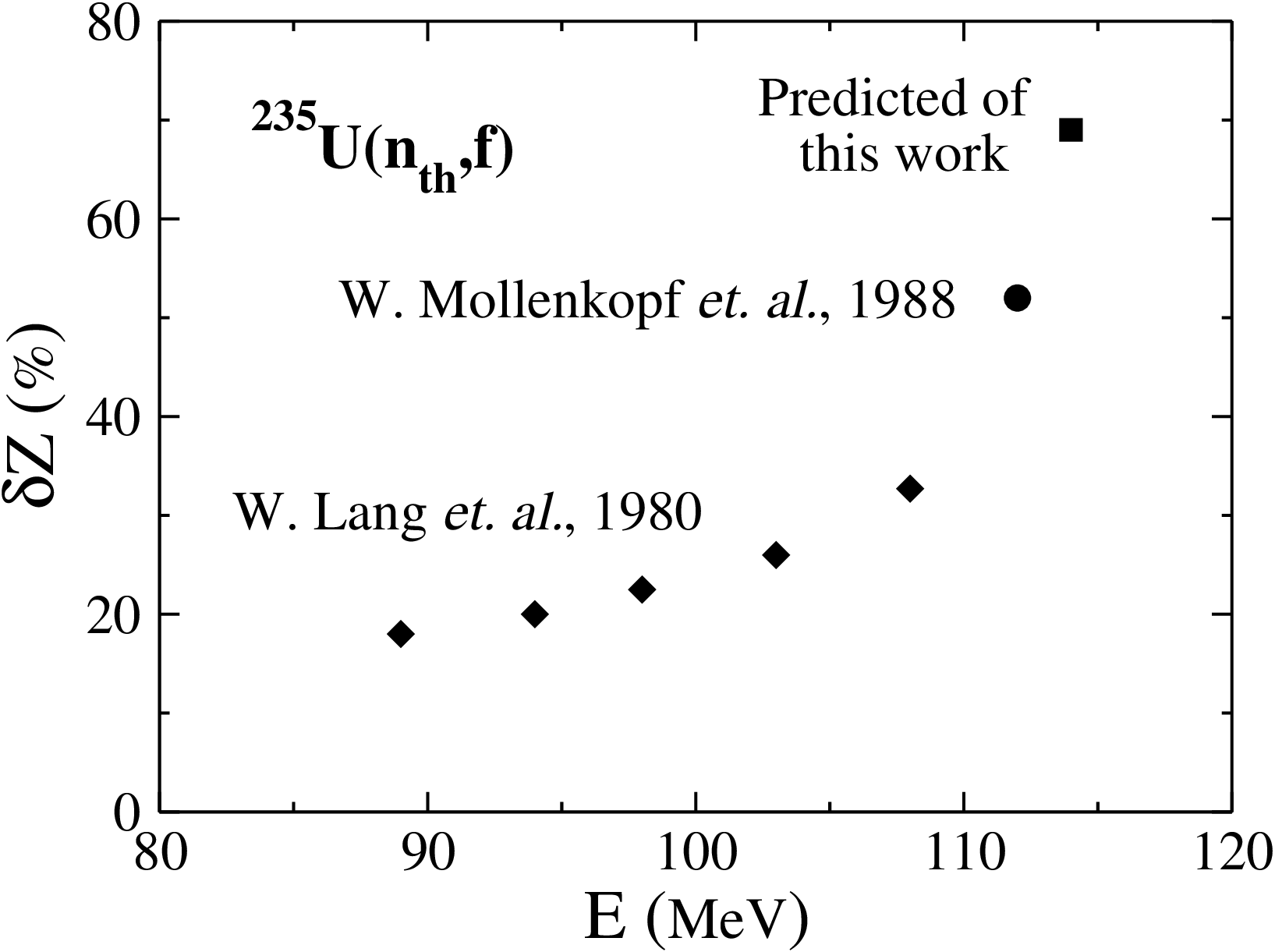}
 \caption{Odd-even effects on the proton distribution as a function of $E$ in $^{233}$U(n$_{th}$,f). Values are taken from Ref. [\onlinecite{Lang80}] and [\onlinecite{Mollenkopf92}], except the higher one, which corresponds to prediction of this work. See text.}
 \label{fig8}
 \end{figure}

\begin{figure}
 \centering
 \includegraphics[width=8cm]{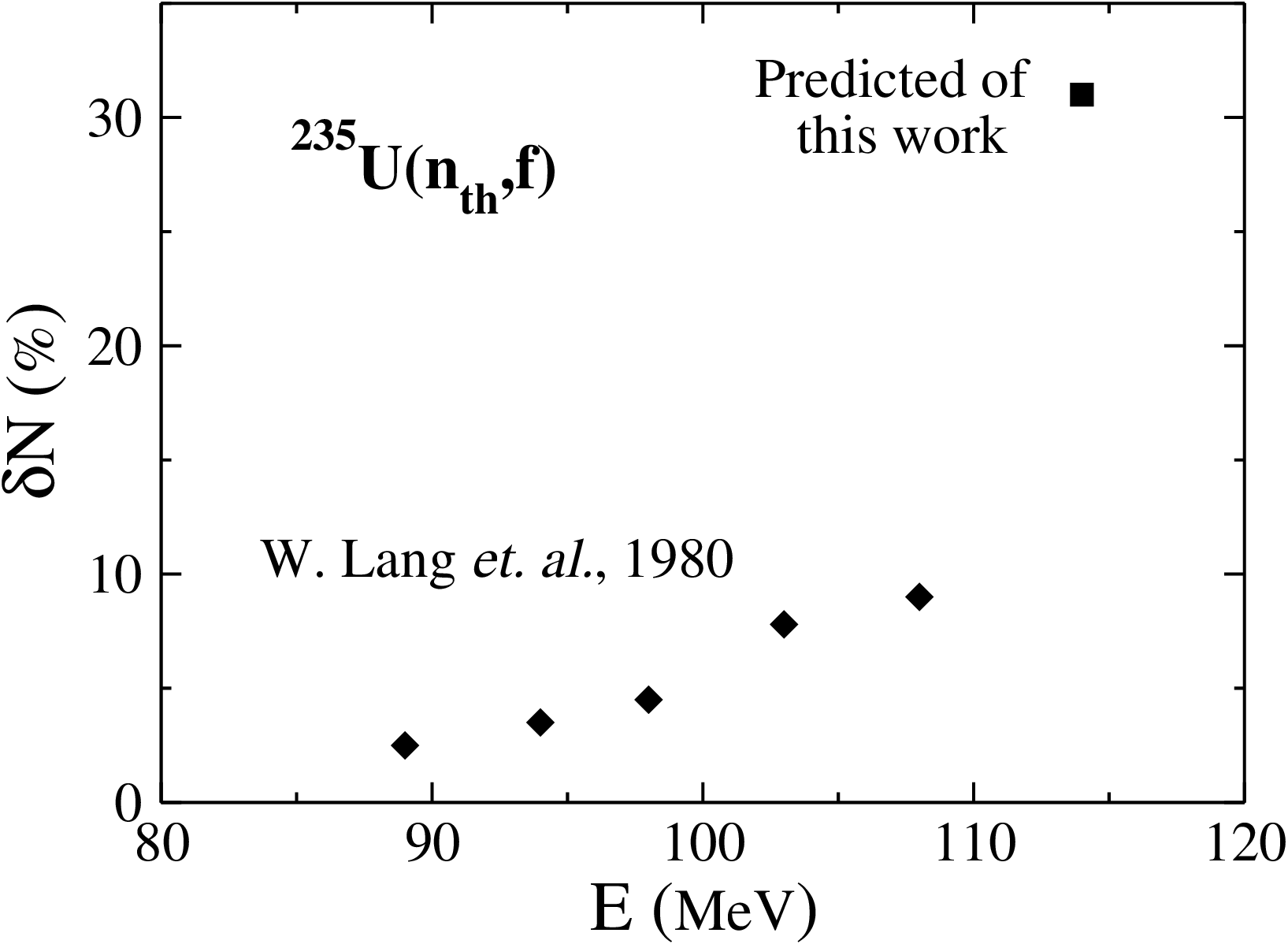}
 \caption{Odd-even effects on the neutron distribution as a function of $E$ in $^{235}$U(n$_{th}$,f). Values are taken from Ref. [\onlinecite{Lang80}], except the higher one, which corresponds to prediction of this work. See text.}
 \label{fig9}
 \end{figure}

\begin{figure}
 \centering
 \includegraphics[width=8cm]{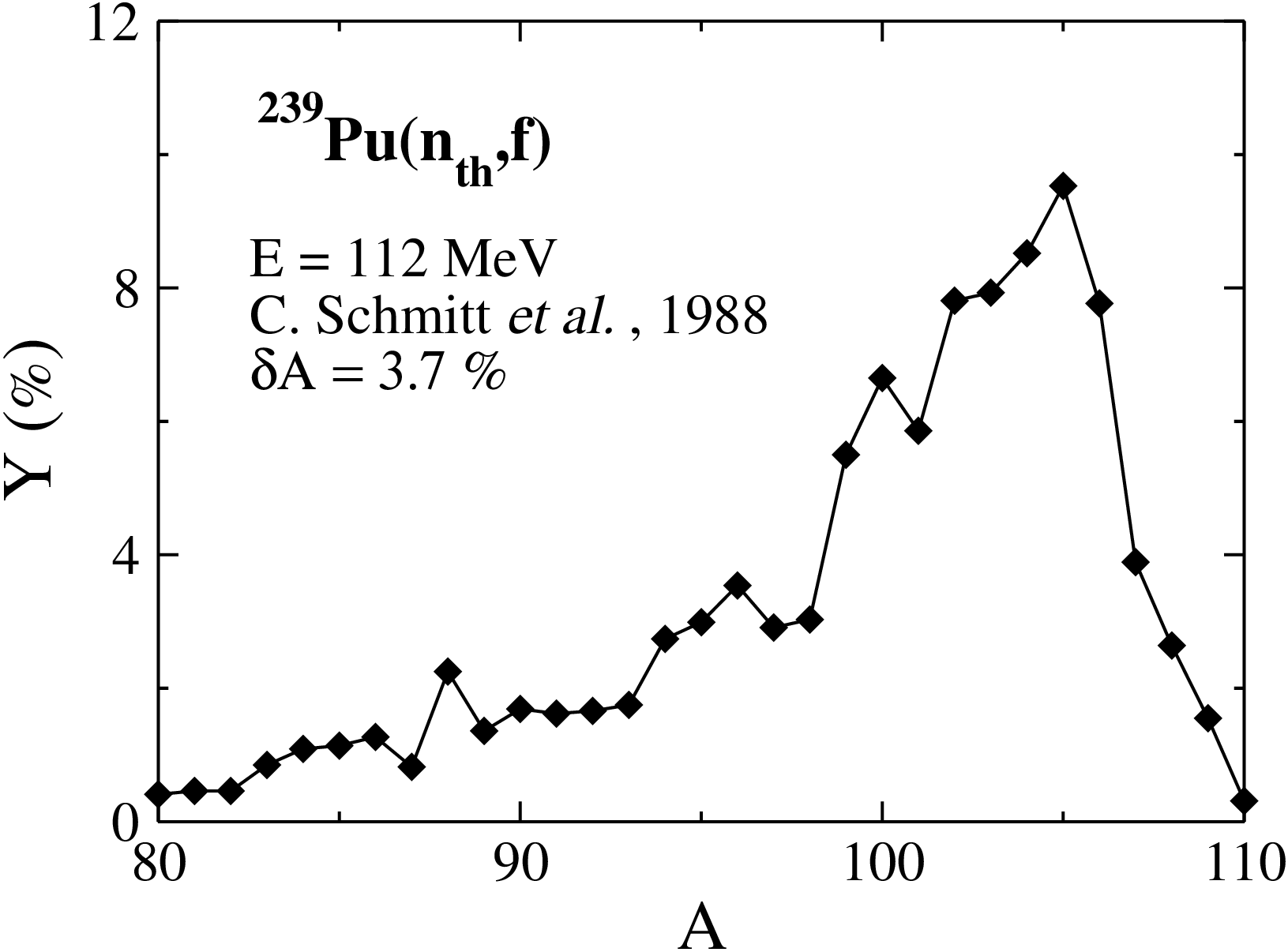}
 \caption{Mass number distribution in the energy window $E=112$ MeV for cold fission from $^{239}$Pu(n$_{th}$,f). Taken from Ref. \cite{Schmitt84}. From these data one calculates $\delta A$= 3.7\%. See text.}
 \label{fig10}
 \end{figure}

\begin{figure}
 \centering
 \includegraphics[width=8cm]{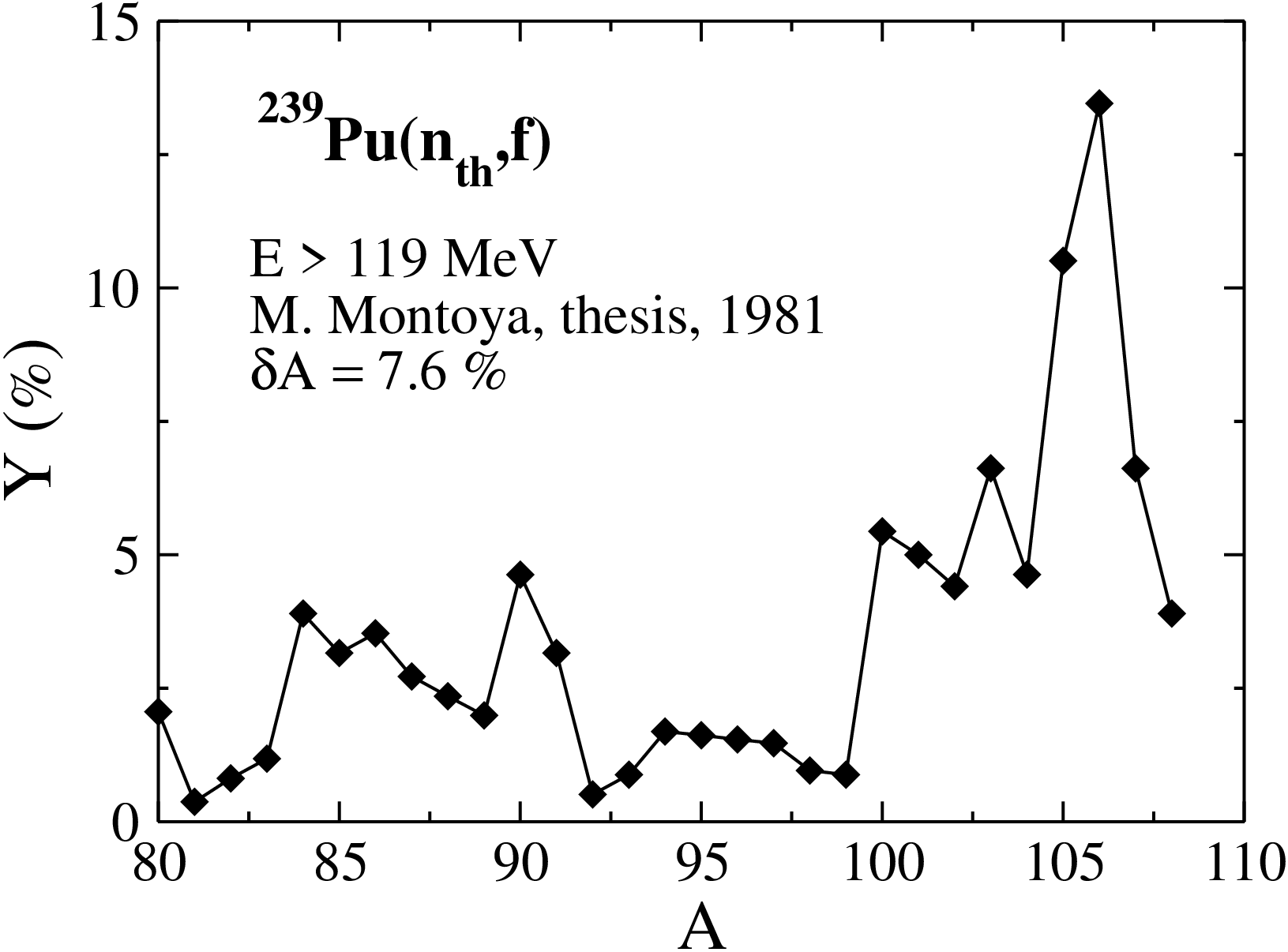}
 \caption{Mass number distribution in the energy window $E>119$ MeV for cold fission from $^{239}$Pu(n$_{th}$,f). Taken from Ref. [\onlinecite{Montoya81}]. From these data one calculates $\delta A$= 7.6\%. See text.}
 \label{fig11}
 \end{figure}

\begin{figure}
 \centering
 \includegraphics[width=8cm]{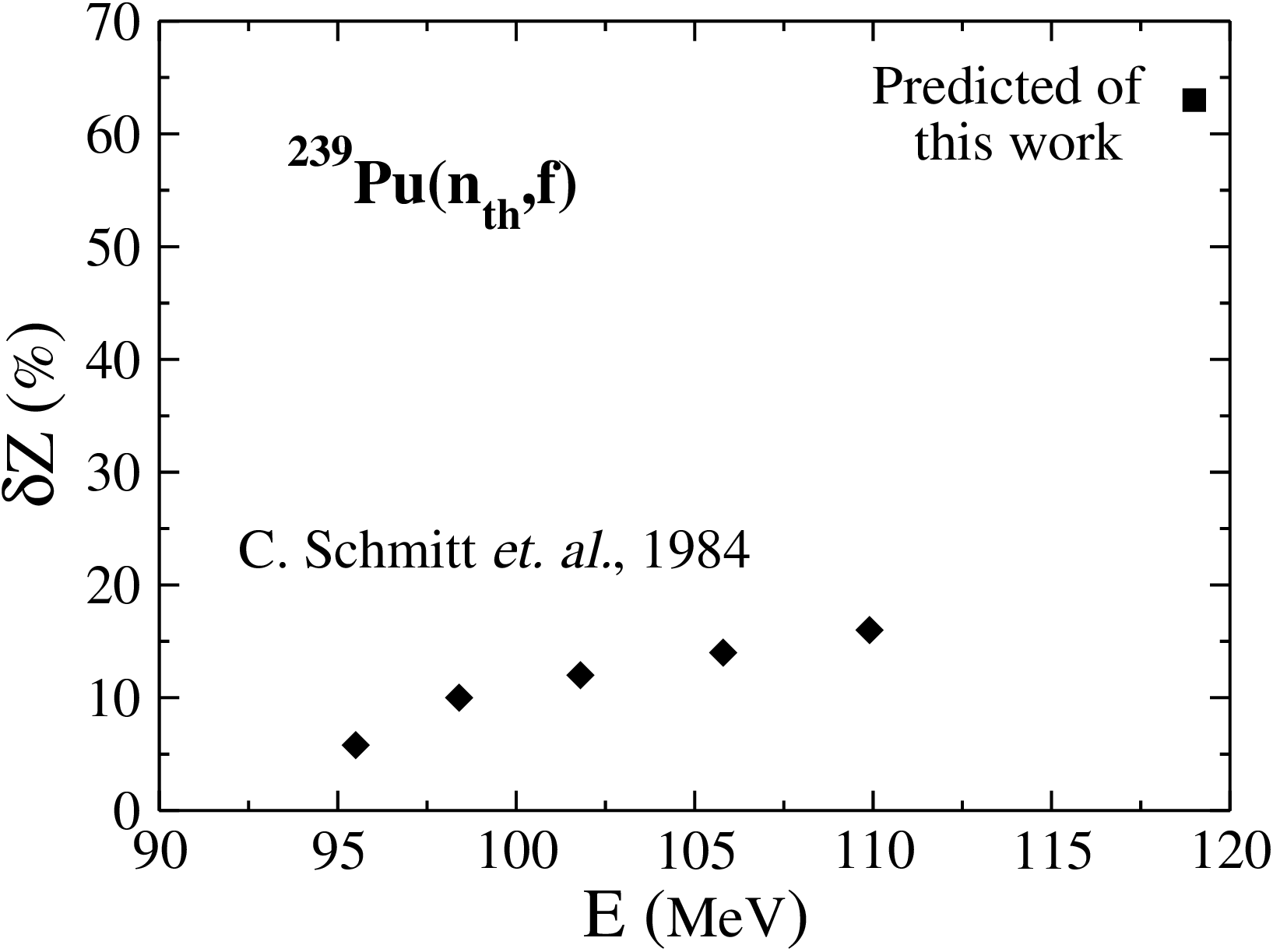}
 \caption{Odd-even effects on the proton distribution as a function of $E$ in $^{239}$Pu(n$_{th}$,f).Values are taken from Ref. [\onlinecite{Schmitt84}], except the higher one, which corresponds to prediction of this work. See text.}
 \label{fig12}
 \end{figure}

\begin{figure}
 \centering
 \includegraphics[width=8cm]{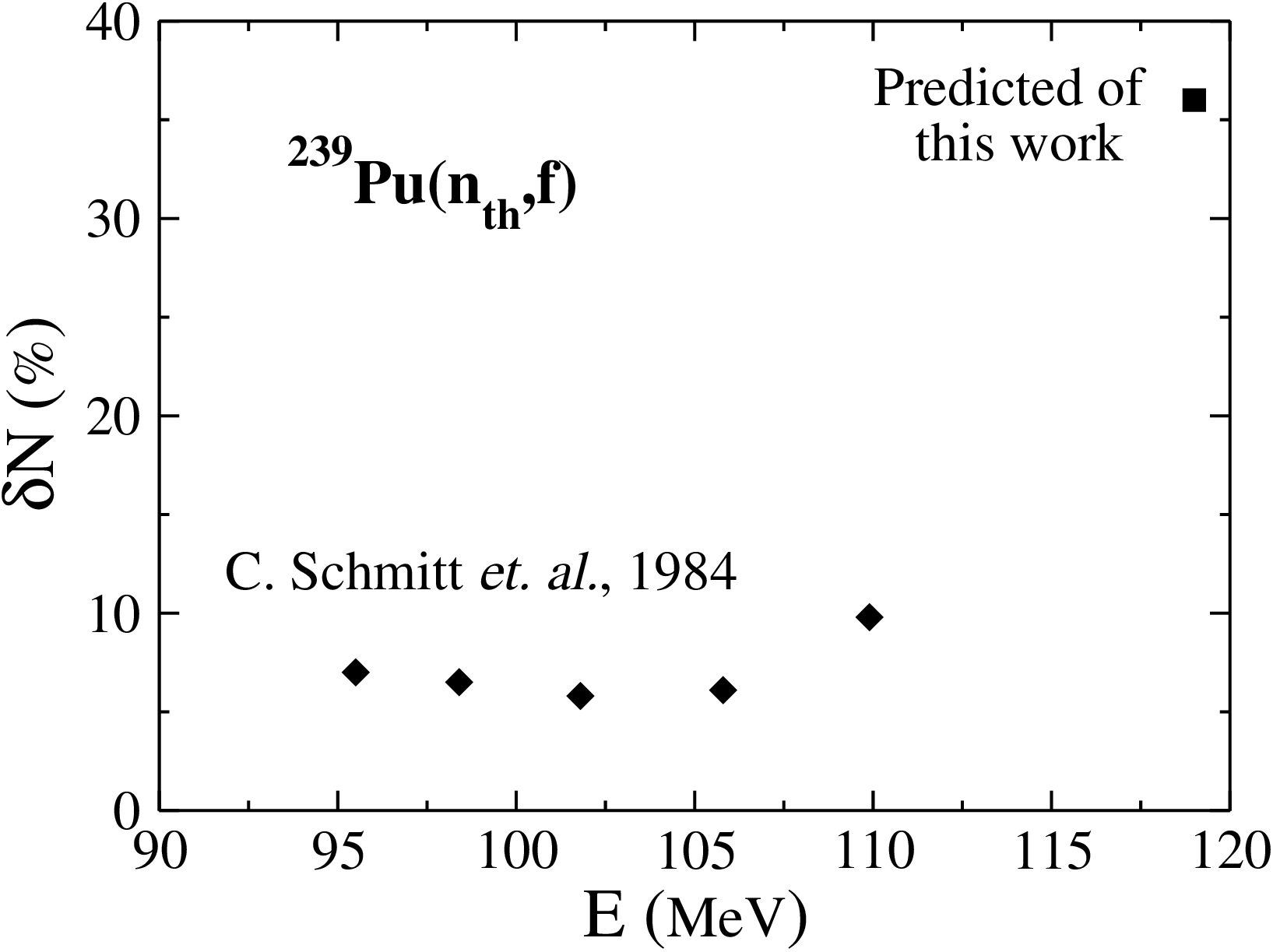}
 \caption{Odd-even effects on the neutron distribution as a function of $E$ in $^{239}$Pu(n$_{th}$,f). Values are taken from Ref. [\onlinecite{Schmitt84}], except the higher one, which corresponds to prediction of this work. See text.}
 \label{fig13}
 \end{figure}

The odd-even effect on the fragment mass number distribution ($\delta A$) is defined by the relation:
\begin{equation*}
 \delta A=\dfrac{Y_{A_{e}}-Y_{A_{o}}}{Y_{A_{e}}+Y_{A_{o}}}
\end{equation*}
where $Y_{A_{e}}$ and $Y_{A_{o}}$ are yields associated to fragments having even and odd mass numbers, respectively.\\\\
Similarly, we define the odd-even effects $\delta N$ and $\delta Z$ on the distribution of neutron number ($N$) and proton number ($Z$), respectively.\\\\
For low energy fission of actinides, the odd-even effect on the proton number distribution of the fragments, previously found to A. L. Wahl \textit{et. al.} \cite{Wahl69} and G. Mariolopoulos \textit{et. al.} \cite{Mariolopoulos81}, suggested that must exist high odd-even effects on the mass number distribution, but they are eroded by neutron emission.\\\\
To study the mass number distributions undisturbed by the emission of neutrons, C. Signarbieux \textit{et. al.} chosen high kinetic energy windows associated to fragments from $^{233}$U(n$_{th}$,f), $^{235}$U(n$_{th}$,f) and $^{239}$Pu(n$_{th}$,f), respectively. The number of events analyzed were 1.5$\times$10$^{6}$, 3$\times$10$^{6}$ and 3.2$\times$10$^{6}$, respectively. These authors used the difference of time of flight technique to separate fragment mass numbers, with solid detectors to measure the fragment kinetic energy. The experiment was conducted in the HFR of ILL in Grenoble \cite{Signarbieux81,Montoya81}.\\\\
Although $\delta A$ is not zero, it is not as high as was expected. However, one may show that this small $\delta A$ is compatible with nucleon pair breaking and higher odd-even effects in the distribution of proton or neutron numbers. \\\\
H.-G. Clerc \textit{et. al.} \cite{Clerc86}, for $E=$ 114.1 MeV, obtained the mass number distribution presented in Fig. \ref{fig2}, from which one calculates $\delta A=$ 30.14 \%. The mass number distribution in the window of light fragment kinetic energy 113.5 $<E<$ 114.5 MeV, in $^{233}$U(n$_{th}$,f), is presented in Fig. \ref{fig3} \cite{Montoya81}. From this distribution one gets $\delta A=$ 32.18 \%.\\\\
In order to interpret values of $\delta A$, $\delta Z$ and $\delta N$, let’s first define, for each $A$, $Z_{0}$ as the light fragment proton number that corresponds to maximal available energy of the reaction ($Q$) Suppose that nucleon pair-breaking occurs, but fragment couple reaching maximum total kinetic energy still corresponding to $Z_0$; therefore, for $^{233}$U(n$_{th}$,f), $\delta Z=$ 66.6 \% and $\delta N=$ 33.3 \%. This result is coherent with the stereographic projection of the values obtained by U. Quade \textit{et. al.} \cite{Quade88} for lower energy windows, as it is presented in Figs. \ref{fig4} and \ref{fig5}.\\\\
For total kinetic energy $TE \geq$ 204 MeV, only $^{104}$Mo ($Z= 42$, $N= 62$) survives \cite{Montoya81}; therefore $\delta A= 1$.\\\\
The mass number distribution for the kinetic energy window 113.5 $<E<$ 114.5 MeV, for light fragments from the reaction $^{235}$U(n$_{th}$,f), is presented in Fig. \ref{fig6}, from which one calculates $\delta A=$ 15.3 \%. For $E=$ 113.7 MeV, W. Lang \textit{et. al.} obtained the distribution presented in Fig. \ref{fig7}, from which one calculates $\delta A \cong$ 23.82 \%. For $E=$ 108 MeV, W. Lang \textit{et. al.} obtain $\delta A \cong$  0, $\delta Z \cong$ 35 \% and $\delta N \cong$ 8 \% \cite{Lang80}. Suppose that proton number correspond to $Z_0$, then one calculates $\delta Z=$ 66.6 \% and $\delta N=$ 33.3 \%. This result is coherent with the stereographic projection of the values obtained by W. Lang \textit{et. al.} \cite{Lang80} for windows of lower energies, as it is shown in Figs. \ref{fig8} and \ref{fig9}. \\\\
For total kinetic energy $TE\geq$ 203 MeV, the surviving mass numbers are $A=$ 104-106, among which there are 16 even and 8 odd mass numbers; therefore $\delta A=$ 30 \%. For $TE\geq$ 200 MeV J. Trochon \textit{et. al.} \cite{Trochon86} measured mass number and proton number distributions. From their results one obtains $\delta A \cong$ 20 \%; $\delta Z \cong$ 60 \%, and $\delta N \cong$ 60 \%.\\\\
The mass number distribution in the window of light fragment kinetic energy $E=$ 112 MeV, in $^{239}$Pu(n$_{th}$,f), obtained by C. Schmitt \textit{et. al.}, is presented in Fig. \ref{fig10} \cite{Schmitt84}. From this data one calculates $\delta A \cong$ 3.7 \%. For this energy they obtain $\delta Z \cong$ 15 \%, and $\delta N \cong$ 10 \% \cite{Schmitt84}. For $E>$ 119 MeV, one obtains the distribution presented in Fig. \ref{fig11}, from which one calculates $\delta A=$ 7.6 \%.
Suppose that for $E>119$ MeV, the proton numbers correspond to $Z_0$, therefore $\delta Z=$ 66.6 \% and $\delta N=$ 33.3 \%, which are coherent with the stereographic projection of the values obtained by C. Schmitt \textit{et. al.}, for lower energy windows. See Figs. \ref{fig12} and \ref{fig13}.\\\\
For $TE>210$ MeV, one obtains masses $104\leq A \leq107$ , with the higher yield corresponding to $A=106$. \\\\
For lower kinetic energy windows, one obtains $\delta A=0$. This result does not imply that $\delta Z$ y $\delta N$ values are zero. M. Montoya \textit{et. al.} using a combinatorial analysis demonstrated \cite{Montoya81} that if there is one and only one nucleon (proton o neutron) pair breaking, then
\begin{equation}
 1+\delta A = \delta N + \delta Z.
\end{equation}
H. Nifenecker \textit{et. al.} \cite{Nifenecker82} developed other combinatorial analysis that produces the same relation.

\section{Shell and Coulomb effects on cold fission}
\begin{figure}
 \centering
 \includegraphics[width=8cm]{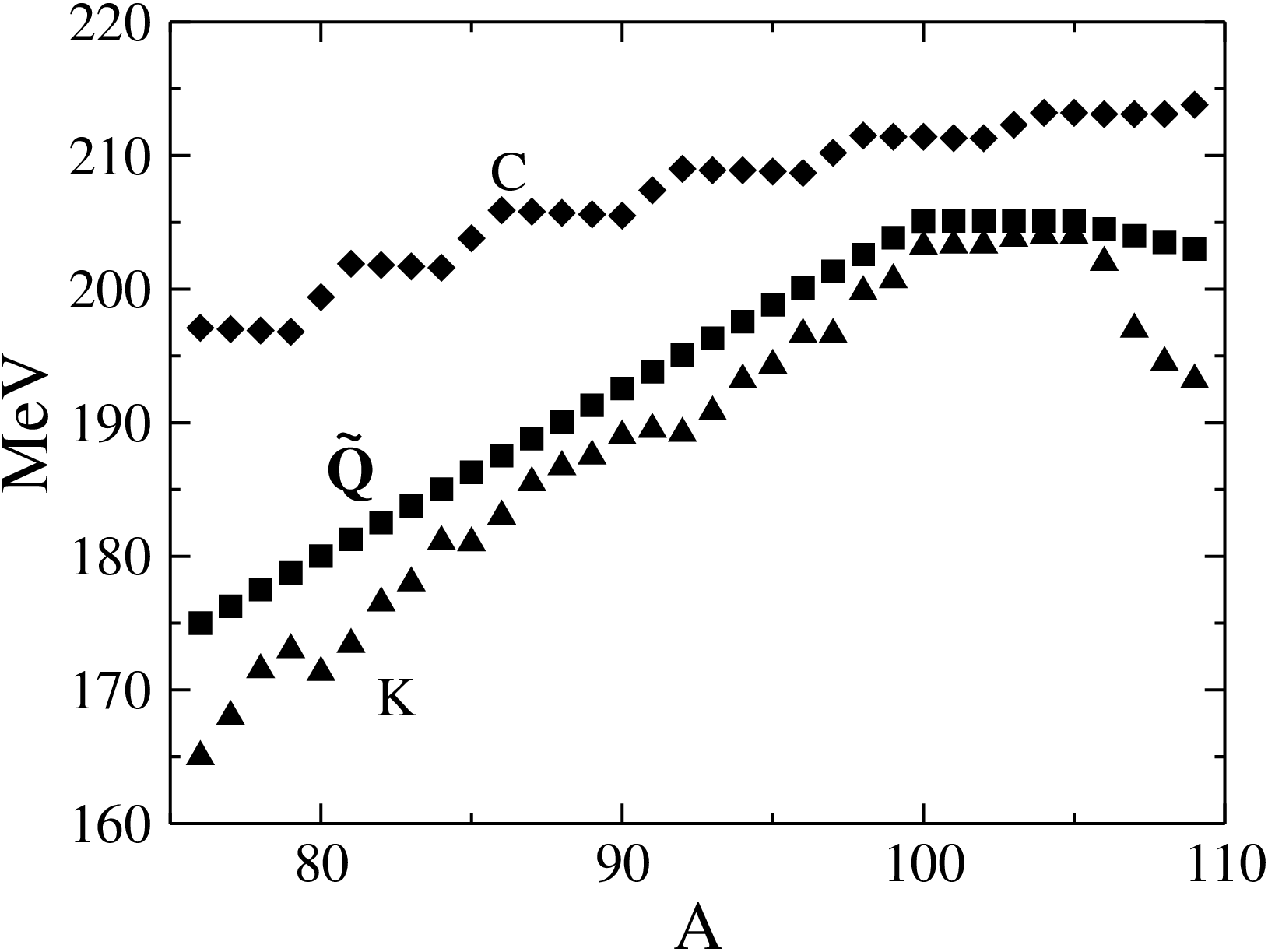}
 \caption{Electrostatic interaction energy between spherical fragments with spherical surfaces separated by 2 fm(C); smoothed curve of maximal available energy ($\tilde{Q}$); and maximal value of the total kinetic energy ($K$) for $^{233}$U(n$_{th}$,f). See text.}
 \label{fig14}
 \end{figure}

\begin{figure}
 \centering
 \includegraphics[width=8cm]{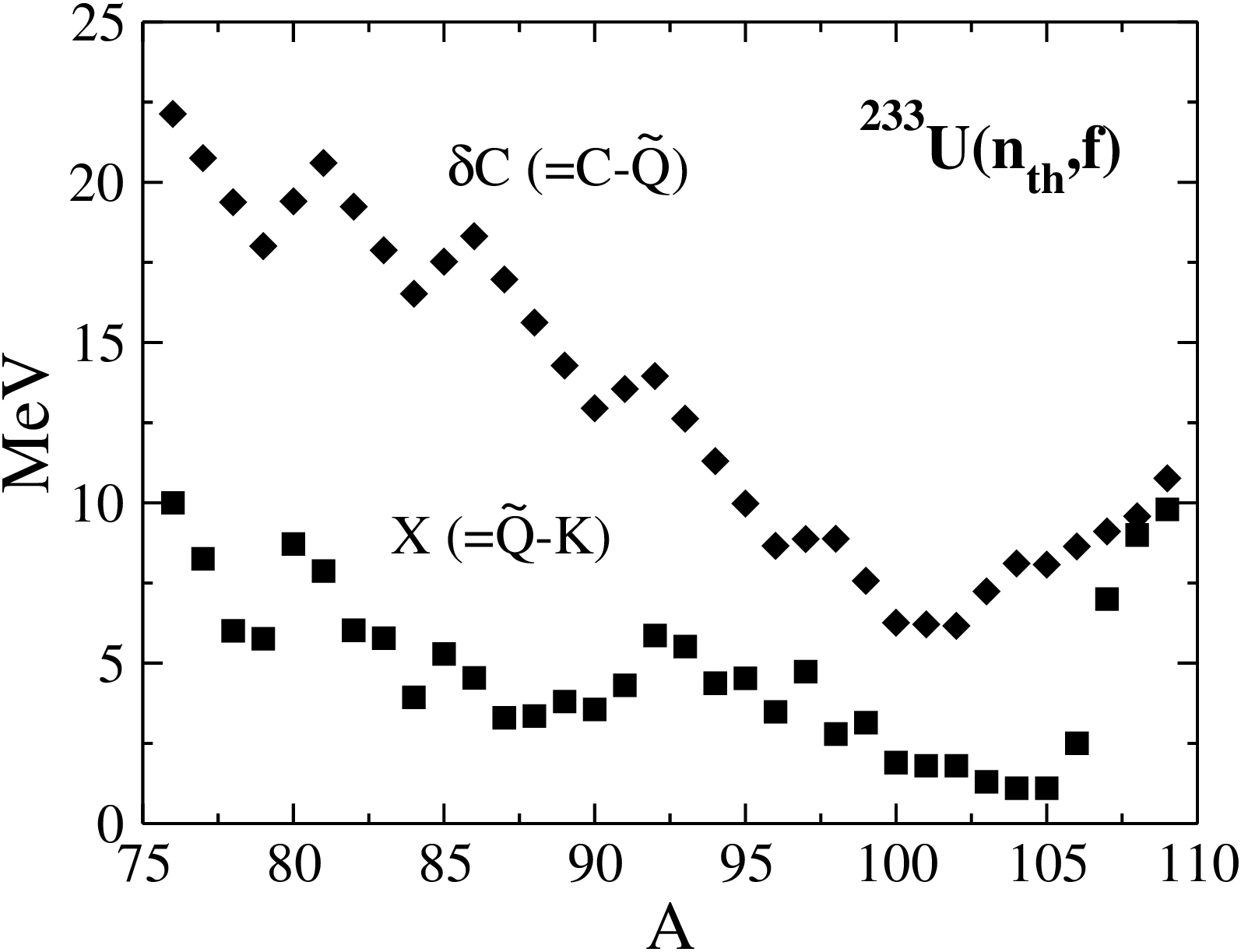}
 \caption{Smoothed curves of electrostatic energy excess $\delta C$ ($=C-\tilde{Q}$) and minimal values of total excitation energy $X$ ($=\tilde{Q}- K$), a function of light fragment mass number $A$ in $^{233}$U(n$_{th}$,f). See text.}
 \label{fig15}
 \end{figure}

\begin{figure}
 \centering
 \includegraphics[width=8cm]{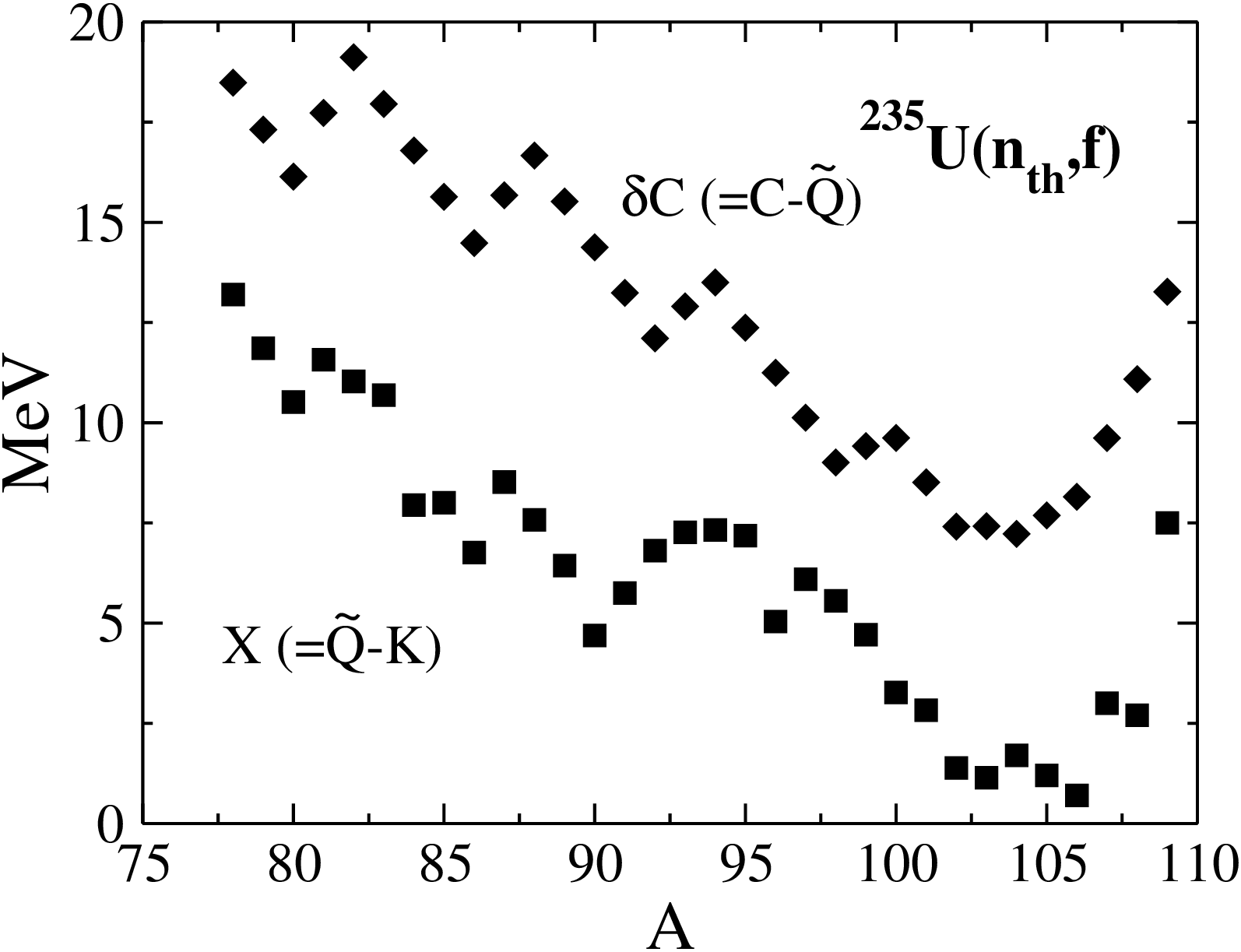}
 \caption{Smoothed curves of electrostatic energy excess $\delta C$ ($=C-\tilde{Q}$) and minimal values of total excitation energy $X$ ($=\tilde{Q}- K$), a function of light fragment mass number $A$ in $^{235}$U(n$_{th}$,f). See text.}
 \label{fig16}
 \end{figure}

\begin{figure}
 \centering
 \includegraphics[width=8cm]{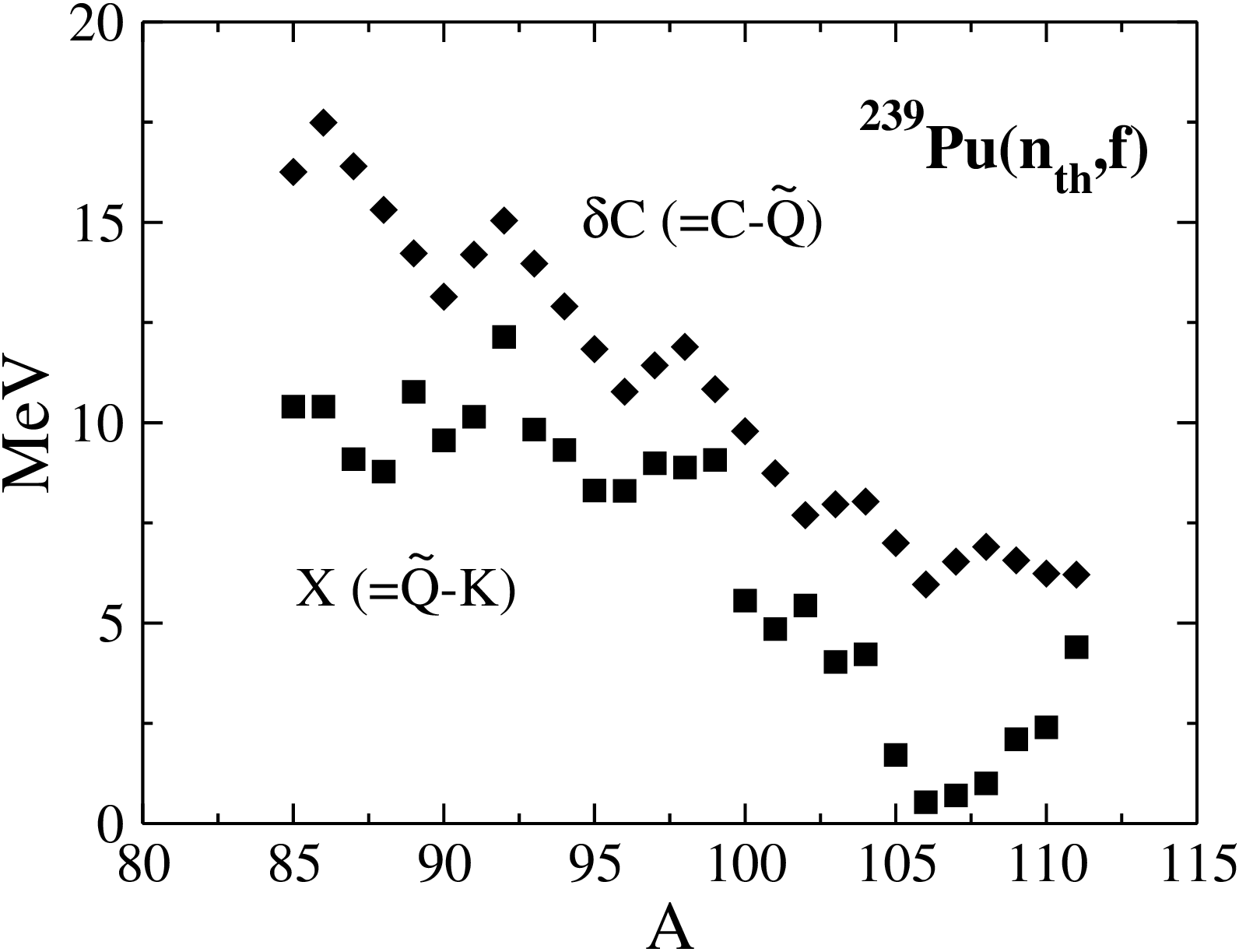}
 \caption{Smoothed curves of electrostatic energy excess$\delta C$ ($=C-\tilde{Q}$) and minimal values of total excitation energy $X$ ($=\tilde{Q}- K$), a function of light fragment mass number $A$ in $^{239}$Pu(n$_{th}$,f). See text.}
 \label{fig17}
 \end{figure}

In cold fission from $^{233}$U(n$_{th}$,f), $^{235}$U(n$_{th}$,f) and $^{239}$Pu(n$_{th}$,f), shell and Coulomb effects on mass number and kinetic distribution were observed \cite{Montoya81}.\\\\
To describe ``Coulomb effect'' let's first define the following curves depending of $A$: i) $\tilde{Q}$, the smoothed curve referred to maximal value of available energy of the reaction ($Q$), as a linear curve in the mass region from the lightest $A$ to that at which ends the linear trend and, for higher mass numbers, a curve following the approximately horizontal trend of $Q$; ii) $C$, the electrostatic interaction energy between complementary spherical fragments corresponding to $Z_0$,with surfaces separated by 2 fm (the result is a step function of mass number); and iii) $K$, the maximal value of total kinetic energy, as the lowest of the ten highest values of total kinetic energy, defined as presented by M. Montoya for $^{233}$U(n$_{th}$,f), $^{235}$U(n$_{th}$,f) and $^{239}$Pu(n$_{th}$,f), respectively \cite{Montoya84}.\\\\
Now let's define the smooth excess of electrostatic energy, as a function of $A$, by the relation
\begin{equation}
 \delta C = C -\tilde{Q}
\end{equation}
and the smoothed values of minimal excitation energy, as a function of $A$, by the relation
\begin{equation}
 X= \tilde{Q}-K
\end{equation}
The curves of $C$, $K$, and $\tilde{Q}$ values for $^{233}$U(n$_{th}$,f) are presented in Fig. \ref{fig14}.\\\\
The fluctuations of $K$ in $^{233}$U(n$_{th}$,f), $^{235}$U(n$_{th}$,f) and $^{239}$Pu(n$_{th}$,f) are interpreted as effects of the step variation of the electrostatic energy as a function of $Z_0$ \cite{Montoya84,Montoya86}.\\\\
It is found that the smoothed values $\delta C$ and $X$ are correlated. This result suggests that, at scission, a higher excess of electrostatic interaction energy produces a higher deformation of fragments, therefore a higher $X$ and $A$ then lower $K$-values.\\\\
For the mass regions $78\leq A \leq100$ in $^{233}$U(n$_{th}$,f); $80\leq A \leq102$ in $^{235}$U(n$_{th}$,f); and $90\leq A\leq108$ in $^{239}$Pu(n$_{th}$,f): $\delta C$ and $X$ are decreasing functions of $A$. See Figs. \ref{fig15}, \ref{fig16} and \ref{fig17}.\\\\
In $^{233}$U(n$_{th}$,f), the higher Q-values, are around 204 MeV, which occurs in the mass region  $100\leq A \leq106$, which corresponds to the lowest values of $\delta C$. See Fig. \ref{fig15}. For complementary fragments ($Z=50$, $N=80$) and ($Z=42$, $N=62$), the $K$ value reach the respective $Q$  value.\\\\
As it is presented in Fig. \ref{fig16}, in the mass region $100\leq A \leq106$, in $^{235}$U(n$_{th}$,f), $\delta C$ well as \textit{X} are very approximately zero.\\\\
In$^{239}$U(n$_{th}$,f), the lowest \textit{X} value occurs in the mass region $100\leq A \leq110$. See Fig. \ref{fig17}. \\\\
For $^{233}$U(n$_{th}$,f), $^{235}$U(n$_{th}$,f) and $^{239}$Pu(n$_{th}$,f), the value of $\delta C$ is not zero. At least one of the fragments must be deformed and spends energy for that. However, molybdenum ($Z= 42$) with $N=60$, $62$ y $64$; and zirconium ($Z= 40$) with $N= 60$, $62$ y $64$; have prolate geometries in their ground states, in addition they are transitional and soft \cite{Montoya81}. At scission, these fragments are deformed without excitation energy, and the electrostatic energy is equal to the $Q$-value. \\\\
For $^{233}$U(n$_{th}$,f), $^{235}$U(n$_{th}$,f) and $^{239}$Pu(n$_{th}$,f), the fluctuations of mass number yield are anti-correlated to $\delta C$ \cite{Wahl69,Mariolopoulos81}.\\\\
For $^{233}$U(n$_{th}$,f), in the light fragment kinetic energy window 113.5 $\leq E \leq$ 114.5 MeV, the mass yield peaks toward a maximum around $A =$ 90 and 100, respectively, which correspond to minimal values of $\delta C$ \cite{Montoya81,Nifenecker82}.
For $^{235}$U(n$_{th}$,f), in the light fragment kinetic energy window 113.5$\leq E \leq$114.5 MeV, the mass yield peaks toward a maximum around $A=$ 90 y 102, respectively, which correspond to minimal values of $\delta C$. For $^{239}$Pu(n$_{th}$,f), in the light fragment kinetic energy region $E>118$ MeV, the mass yield peaks toward a maximum around $A=$ 84, 89, 96, 100 and 106, respectively, which correspond to minimal values of $\delta C$.\\\\
For  $^{233}$U(n$_{th}$,f) as well as for $^{235}$U(n$_{th}$,f), for$ E>106$, $\delta C$ increases sharply with A and, therefore, $K$ decreases sharply too. For the reaction $^{235}$U(n$_{th}$,f), $\delta C$ decrease for $A>110$.\\\\
The fluctuations of $X$ and $\delta C$ are correlated and they increase with asymmetry of fragmentations. This is because the step-variation of the electrostatic interaction energy between fragments produced, by changing a unit of proton number, is more pronounced in asymmetric region, as one may in Fig. \ref{fig17}, corresponding to $^{233}$U(n$_{th}$,f).\\\\
The electrostatic interaction energy between complementary fragments constitutes a barrier to fission in the channel of the corresponding charge and mass. The probability to produce a couple of fragments will be higher for lower difference between this barrier and the available energy. Thus, for a same $Q$-value, between neighbor mass numbers, the fragments with higher asymmetry of charge will reach higher values of kinetic energy. This is equivalent to say for $TE$ close to the maximal value the prevalent fragmentations will be those with higher asymmetry. The Coulomb effects in cold fission will be present also in proton number yields corresponding to odd proton number. The higher yield corresponds to higher asymmetry, as it was shown by W. Schwab \textit{et. al.} in the case of $^{233}$U(n$_{th}$,f) \cite{Schwab94}.

\section{Conclusions}
 Neutron emission has erosive consequences on fragment mass and kinetic energy distribution in $^{233}$U(n$_{th}$,f), $^{235}$U(n$_{th}$,f) and $^{239}$Pu(n$_{th}$,f). To avoid those disturbances, the cold fission, i.e. very low excitation region, was studied. Contrary to expected, in cold fission, nucleon pair breaking still exists. Nevertheless at the maximal total kinetic energy only even-even fragments survive. For lower energies, zero odd-even effect on mass distribution does not exclude non zero odd-even effects on proton or neutron distributions. \\\\
 In addition, in cold fission, Coulomb effects on charge, mass and kinetic energy distribution are present: for two fragmentations with equal $Q$-value, the higher yield corresponds to the higher charge asymmetry. Moreover, Coulomb effects are presented as fluctuations of the maximal kinetic energy value as a function of mass. This effect is due to the fact that, each five units of mass number, there is a change of the proton number that maximizes the $Q$-value, which produces a step variation of electrostatic interaction energy between complementary fragments with the same shapes. A similar effect on mass number yield curve is observed. Due to the characteristics of electrostatic interaction, those effects are more notorious in higher asymmetry region. Finally, results from several experiments in cold fission suggest that scission explore all configurations permitted by the $Q$-value.
%
%
%

%

%


\end{document}